# Photoinduced vibronic coupling in two-level dissipative systems

A. V. Ivanov,*
*Research Center "Information Optical Technologies",*
*ITMO University, Birzhevaya liniya, 14, Saint Petersburg, 199034, Russia*

Interaction of an electron system with a strong electromagnetic wave leads to rearrangement both the electron and vibrational energy spectra of a dissipative system. For instance, the optically coupled electron levels become split in the conditions of the ac Stark effect that gives rise to appearance of the nonadiabatic coupling between the electron and vibrational motions. The nonadiabatic coupling exerts a substantial impact on the electron and phonon dynamics and must be taken into account to determine the system wave functions. In this paper, the vibronic coupling induced by the ac Stark effect is considered. It is shown that the interaction between the electron states dressed by an electromagnetic field and the forced vibrations of reservoir oscillators under the action of rapid changing of the electron density with the Rabi frequency is responsible for establishment of the photoinduced vibronic coupling. However, if the resonance conditions for the optical phonon frequency and the transition frequency of electrons in the dressed state basis are satisfied, the vibronic coupling is due to the electron-phonon interaction. Additionally, photoinduced vibronic coupling results in appearance of the doubly dressed states which are formed by both the electron-photon and electron-vibrational interactions.

## I. INTRODUCTION

The optical properties of some solid state systems are substantially determined by vibronic coupling (VC) [1]. Among these systems are molecular crystals [2], semiconductor quantum dots (QDs) [3], rare-earth (RE) ion doped crystals [4], and nitrogen-vacancy (NV) centers in diamond [5]. The listed materials have rich vibrational spectra associated with the molecular dynamics in a dissipative medium. However, the strength of the VC for these materials is different. The VC in molecular crystals and QDs is relatively weak while the materials with molecules formed by the crystal lattice defects (RE ions, and NV centers in our list) exhibit the especially strong VC. In the latter case, the strong VC is provided by the nonadiabatic interaction between the localized vibrational modes and electrons [6, 7].

Interaction of a dissipative system with a strong electromagnetic field may lead to enhancement or, even, to appearance of the VC which is absent without a field. Such situation is present due to changing the energy spectrum of both the electron [8, 9] and vibrational states [10] in the field of a strong electromagnetic wave. In the case of the ac Stark effect, electron levels of the system become split with the energy gaps which are comparable with the energies of vibrational modes. Therefore, the electron and vibrational motions in a dissipative system cannot be separated, and the nonadiabaticity of coupling must be taken into account [11]. At the same time, the Rabi oscillations of the electron density lead to the periodic displacement of equilibrium positions of reservoir oscillators [12, 13] that means appearance of an additional mode in the vibrational spectrum. It is worth to note that formation of the vibronic states in the conditions of the ac Stark effect can be interpreted as a manifestation of the pseudo-Jahn-Teller effect [14] in the electron-photon system. In this case, a strong electromagnetic field both removes the degeneracy that exists in the system consisting of the electron and photon subsystems and creates a new basis of states, so-called dressed states (DSs), which have closely spaced energies. Since the reservoir vibrational modes are able to mix the closely spaced DSs, the photoinduced VC (PIVC) can be established. The dynamical enhancement of the VC in the InGaAs QDs in the presence of an intense optical pulse was observed in recent experiments [15]. It was shown in Ref. [15] that at low light intensities the electron-phonon coupling in QDs was relatively weak while at high light intensities the coupling with acoustic phonons was enhanced.

In this paper, we consider the PIVC in two-level dissipative systems (TLDS) taking into account different vibrational modes. In Sec. II, we propose a physical mechanism of formation of the VC with use of DS picture. The mechanism is based on appearance of the forced vibrations of reservoir oscillators under the action of rapid changing of the electron density with the Rabi frequency and back action of the photoinduced vibrations on the electron density. Then, contributions of the different vibrational modes into the PIVC are discussed. In Sec. III, we present the photoinduced vibronic states (PIVSs) as the doubly dressed states (DDSs) formed by the interaction of electrons with the optical phonon modes or the forced vibrations of reservoir oscillators.

## II. MECHANISM OF VIBRONIC COUPLING FORMATION

A two-level electron subsystem interacting with a strong monochromatic electromagnetic field and a phonon reservoir is considered. The Hamiltonian of the system can be written as

$$H = H_1 + H_2, \quad H_1 = \frac{\varepsilon}{2}\sigma_z + \hbar\omega_0 c^+ c + \hbar g_0 \sigma_+ c + \hbar g_0^* c^+ \sigma_-,$$
$$H_2 = H_{pn}^0 + V_{e\text{-}pn}. \quad (1)$$

Here, $\varepsilon$ is the energy difference between the ground $|\varphi_1\rangle$ and excited $|\varphi_2\rangle$ electron levels; $c$ and $c^+$ are the annihilation and creation operators of the photon of frequency $\omega_0$; $g_0$ is the coupling strength of the electron-photon interaction; $\sigma_z$, $\sigma_+$, and $\sigma_-$ are the Pauli operators; $H_{pn}^0$ is the Hamiltonian of the non-interacting phonon subsystem; $V_{e\text{-}pn}$ is the electron-vibrational interaction Hamiltonian. In Eq. (1) we divided the total Hamiltonian into two parts in order to include the electron-photon interaction into the zeroth-approximation. The first part contains the electron and photon subsystems and their interaction in the dipole approximation and rotating wave approximation (RWA) and the second part involves the operators related to the vibrational subsystem.

The solution of the Schrödinger equation for the Hamiltonian $H_1$ can be written in the basis of the DSs. It is well known that a form of the solution depends on initial populations of the states. If initially only the ground electron state is populated as will be assumed in what follows, the solution has the form [16]

$$|\Psi(t),m,n\rangle = C_m U_m^+ |\xi_1(t),m,n\rangle + C_m S_m U_m^- |\xi_2(t),m,n\rangle,$$
$$U_m^\pm = \sqrt{\frac{1}{2}\left(1\pm\frac{\delta}{2\Omega}\right)}, \quad S_m = \frac{g_0}{|g_0|}, \quad 2\Omega = \sqrt{\delta^2 + 4|g_0|^2(m+1)},$$
$$\delta = \hbar^{-1}\varepsilon - \omega_0. \quad (2)$$

Here, $C_m$ is the initial probability amplitude for the state with $m$ photons; $n$ is the number of phonons; $\Omega$ is the off-resonance Rabi frequency; $\delta$ is the detuning of the optical field with respect to the electron transition frequency. It should be stressed that the wave function of Eq. (2) is defined for the instant start of the electron-photon interaction in the so-called diabatic case [16]. We present the DSs in the following form for convenience of further consideration

$$|\xi_1(t),m,n\rangle = U_m^+ e^{i\left(\Omega-\frac{\delta}{2}\right)t}|\varphi_1(t),m+1,n\rangle$$
$$-S_m U_m^- e^{i\left(\Omega+\frac{\delta}{2}\right)t}|\varphi_2(t),m,n\rangle,$$
$$|\xi_2(t),m,n\rangle = S_m^* U_m^- e^{-i\left(\Omega+\frac{\delta}{2}\right)t}|\varphi_1(t),m+1,n\rangle$$
$$+U_m^+ e^{-i\left(\Omega-\frac{\delta}{2}\right)t}|\varphi_2(t),m,n\rangle. \quad (3)$$

Then, including $H_1$ into the zeroth-approximation Hamiltonian of $H_2$, we define the matrix element of the electron-vibrational interaction operator as

$$H' = \langle\Psi(t),m,n|\tilde{V}_{e\text{-}pn}|\Psi(t),m,n\rangle,$$
$$\tilde{V}_{e\text{-}pn} = \exp(-i\hbar^{-1}H_0 t)V_{e\text{-}pn}\exp(i\hbar^{-1}H_0 t),$$
$$H_0 = H_1 + H_{pn}^0. \quad (4)$$

Eq. (4) determines the electron-vibrational interaction in the electron-photon system and can be written in two different bases, in the basis of the bare states (BSs) and in the basis of the DSs [see Eq. (2) and (3)]. The energy separation between the DSs, $2\hbar\Omega$, is comparable with the vibrational energy of reservoir modes and, therefore, the DSs can be mixed by the electron-vibrational interaction. By contrast, the BSs are mixed by the electron-photon interaction that gives rise to the electron density oscillations with the Rabi frequency. The oscillations of the electron density initiate the forced vibrations of reservoir oscillators between equilibrium positions of the ground and excited BSs. In order to take into account the effect of displacement of the equilibrium positions (EDEP) of vibrational coordinates, we leave in Eq. (4) only the non-diagonal matrix elements in the DS basis and write these elements using relationships of Eq. (3). Thus, the non-diagonal part of Eq. (4) in the basis of the DSs takes the form

$$C_m^2 S_m U_m^- U_m^+ \langle\xi_1(t),m,n|\tilde{V}_{e\text{-}pn}|\xi_2(t),m,n\rangle + \text{H.c.}$$
$$= \left(C_m U_m^+ U_m^-\right)^2 \left(\tilde{V}_{11} - \tilde{V}_{22}\right)e^{-i2\Omega t}$$
$$+ C_m^2 U_m^- U_m^+ \left(S_m (U_m^+)^2 e^{-i(2\Omega-\delta)t}\tilde{V}_{12}\right.$$
$$\left. - S_m^* (U_m^-)^2 e^{-i(2\Omega+\delta)t}\tilde{V}_{21}\right) + \text{H.c.},$$
$$\tilde{V}_{ij} = \langle\varphi_i(t),m,n|\tilde{V}_{e\text{-}pn}|\varphi_j(t),m,n\rangle. \quad (5)$$

The next step of our consideration is to define the matrix elements $\tilde{V}_{ij}$ in the basis of the BSs taking into account the EDEP. In our consideration, we assume that the TLDS state can be presented as a product state which is composed by its subsystem states in the zeroth-approximation of the electron-vibrational interaction. At the same time, it is worth to note that the phonon functions associated with the ground and excited electron states are different and not orthogonal. Orthogonality of the system states exists due to the electron wave functions. Thus, the photoinduced displacement of equilibrium positions of vibrational coordinates gives rise to renormalization of the phonon frequencies. Since this frequency effect does not participate in the PIVC formation, we will neglect it in the further consideration. Then, to take into account the EDEP, we will discriminate the non-interacting phonon Hamiltonians related to the ground and excited BSs. Therefore, the displacement operator appears only in the phonon Hamiltonian of the excited state.

In the first order with respect to the small displacements, we introduce the operator [17]

$$\Delta H = \left(H_{pn}^0\right)_2 - \left(H_{pn}^0\right)_1 = -\sum_{\mathbf{k},s}\hbar^2\omega_{\mathbf{k},s}f_{\mathbf{k},s}\left(\tilde{b}_{\mathbf{k},s}+\tilde{b}_{-\mathbf{k},s}^+\right),$$

$$f_{\mathbf{k},s} = \sum_{l,\alpha}\frac{\Delta_{l,\alpha}\cdot\mathbf{e}_{\mathbf{k},s}^\alpha}{\sqrt{2Nm_\alpha\hbar\omega_{\mathbf{k},s}}}e^{i\mathbf{k}\cdot\mathbf{R}_{l,\alpha}^g}, \quad \Delta_{l,\alpha} = \mathbf{R}_{l,\alpha}^e - \mathbf{R}_{l,\alpha}^g. \quad (6)$$

Here, $b_{\mathbf{k},s}$ and $b_{\mathbf{k},s}^+$ are the phonon annihilation and creation operators of mode $s$, wave vector $\mathbf{k}$, and frequency $\omega_{\mathbf{k},s}$; $\mathbf{e}_{\mathbf{k},s}^\alpha$ are the normalized polarization vectors of vibrational coordinates for an $\alpha$-type oscillator of mass $m_\alpha$; $N$ is the number of reservoir oscillators; $\Delta_{l,\alpha}$ is the displacement operator related to the $\alpha$-type oscillator position labeled by $l$; $\mathbf{R}^g$ and $\mathbf{R}^e$ are the equilibrium positions of oscillations of the normal coordinates associated with the ground and excited BSs, respectively.

Then, we present the electron-vibrational interaction operator as an expansion of the Taylor's series in the displacements in the following form

$$V_{e\text{-pn}} = \sum_{l,\alpha}\left\{V\left[\mathbf{r}-\mathbf{R}_{l,\alpha}^g\right]+\left(\mathbf{R}_{l,\alpha}-\mathbf{R}_{l,\alpha}^g\right)\frac{\partial V\left[\mathbf{r}-\mathbf{R}_{l,\alpha}^g\right]}{\partial\mathbf{R}_{l,\alpha}}\right.$$
$$\left.+\frac{\left(\mathbf{R}_{l,\alpha}-\mathbf{R}_{l,\alpha}^g\right)^2}{2}\frac{\partial^2 V\left[\mathbf{r}-\mathbf{R}_{l,\alpha}^g\right]}{\partial\mathbf{R}_{l,\alpha}^2}+\ldots\right\}. \quad (7)$$

We consider the second term which has the first order with respect to the displacements. Taking use of the Fourier series decomposition $V(\mathbf{r}) = N^{-1}\sum_{\mathbf{k}}v(\mathbf{k})e^{i\mathbf{k}\cdot\mathbf{r}}$ and presenting the displacements via second quantization operators

$$\mathbf{R}_{l,\alpha} - \mathbf{R}_{l,\alpha}^g = \sum_{\mathbf{k},s,\alpha}\sqrt{\frac{\hbar}{2Nm_\alpha\omega_{\mathbf{k},s}}}\mathbf{e}_{\mathbf{k},s}^\alpha e^{i\mathbf{k}\cdot\mathbf{R}_{l,\alpha}^g}\left(b_{\mathbf{k},s}+b_{-\mathbf{k},s}^+\right), \quad (8)$$

we can write the matrix element related to the ground state in the usual form

$$\tilde{V}_{11}^1 = \tilde{V}_{11}',$$
$$\tilde{V}_{lj}' = \sum_{\mathbf{k},s,n}\hbar g_{lj}^{\mathbf{k},s}e^{(l-j)i\delta t}\left(\sqrt{n_{\mathbf{k},s}}e^{-i\omega_{\mathbf{k},s}t}+\sqrt{n_{\mathbf{k},s}+1}\,e^{i\omega_{\mathbf{k},s}t}\right),$$
$$g_{ij}^{\mathbf{k},s} = \sum_\alpha\frac{-i\mathbf{k}\cdot\mathbf{e}_{\mathbf{k},s}^\alpha v(\mathbf{k})}{\sqrt{2Nm_\alpha\hbar\omega_{\mathbf{k},s}}}\langle\varphi_i|e^{i\mathbf{k}\cdot\mathbf{r}}|\varphi_j\rangle. \quad (9)$$

Here, $g_{ij}^{\mathbf{k},s}$ is the coupling strength of the electron-phonon interaction; $n_{\mathbf{k},s}$ is the phonon number of mode $s$ and wave vector $\mathbf{k}$; $\mathbf{r}$ is the position operator; $v(\mathbf{k})$ is the Fourier transform of the electron-vibrational interaction operator.

To obtain the matrix element of the excited state, we make use of the Feynman formula for the "disentangling" of the exponential operator factor [18]

$$\exp\left\{-i\hbar^{-1}\left(H_{pn}^0\right)_2 t\right\} = \exp\left\{-i\hbar^{-1}\left[\Delta H+\left(H_{pn}^0\right)_1\right]t\right\}$$
$$= \exp\left\{-i\hbar^{-1}\left(H_{pn}^0\right)_1 t\right\}\mathrm{T}\exp\left\{-i\hbar^{-1}\int dt'\Delta H(t')\right\},$$
$$\Delta H(t) = \exp\left\{-i\hbar^{-1}\left(H_{pn}^0\right)_1 t\right\}V_{e\text{-pn}}\exp\left\{i\hbar^{-1}\left(H_{pn}^0\right)_1 t\right\}. \quad (10)$$

Thus, up to the first order with respect to the displacements, the remaining matrix elements have the following form (for details see Appendix A)

$$\tilde{V}_{lj}^1 = \left(\tilde{V}_{lj}'+\Delta\tilde{V}_{lj}'\right)e^{(l-j)i\delta t},$$
$$\Delta\tilde{V}_{lj}' = \sum_{\mathbf{k},s}(l+j-2)\hbar^2 g_{lj}^{\mathbf{k},s}f_{\mathbf{k},s}. \quad (11)$$

Here, $\Delta\tilde{V}_{mn}'$ are the terms that describe the contribution of the EDEP to the electron-vibrational interaction and $\tilde{V}_{mn}'$ are the terms that define the usual electron-phonon interaction [see Eq. (9)]. Consequently, the non-diagonal part of the electron-vibrational interaction in the DS basis can be represented as a sum of these two contributions. Substituting the expressions for the matrix elements of Eqs. (9) and (11) into Eq. (5), we obtain the EDEP contribution as

$$H_{ED}' = \hbar F_0 e^{-i2\Omega t}+\hbar F_0^* e^{i2\Omega t},$$
$$F_0 = \sum_{\mathbf{k},s}\hbar f_{\mathbf{k},s}U_m^+ U_m^- C_m^2\left[(U_m^+)^2 S_m g_{12}^{\mathbf{k},s}-(U_m^-)^2 S_m^* g_{21}^{\mathbf{k},s}\right.$$
$$\left.-2U_m^+ U_m^- g_{22}^{\mathbf{k},s}\right]. \quad (12)$$

The obtained matrix element describes the vibrations of frequency $2\Omega$ under the action of the Rabi oscillations of the electron density between the BSs. The amplitude of the vibrations depends on the module of the displacement (factor $f_{\mathbf{k},s}$), electron-phonon and electron-photon (factor $U_m^+ U_m^-$) coupling constants. Since the frequency of the vibrations is equal to the frequency of the electron transition between the DSs, these vibrations mix the DSs what stipulates the back action of the vibrations on the electron density. Thus, the interaction of Eq. (12) gives rise to appearance of the PIVC in the TLDS. At the same time, the electron-phonon interaction can also participate in establishment of the PIVC. Let us estimate the EDEP and phonon contributions to the non-diagonal part of the electron-vibrational interaction in the DS basis. Since the frequency of the photoinduced vibrations is resonant to the frequency of the DS transition, all of the reservoir oscillators irradiated by light make contribution in the sum over a wave vector [see Eq. (11)]. By contrast, the main contribution to the VC is made by the acoustic phonon mode of a certain wave vector according to the resonance conditions. Therefore, the ratio of the acoustic phonon and

EDEP contributions is proportional to $N^{-1/2}$. In the case of optical phonon modes, the ratio is inverse because these kinds of vibrations have a weak dependence of the wave vector. In this estimations, we use the fact that $g_{ij}^{\mathbf{k},s} \sim N^{-1/2}$ and $f_{\mathbf{k},s} \sim N^{-1/2}$. However, to form the PIVC with use of an optical phonon mode, we must fulfil the exact resonance conditions. Thus, the mechanism of formation of the PIVC in the TLDS depends on the energy domain of phonon modes. If the energy separation between the DSs is comparable with the energy quantum of an optical mode, the formation of PIVC is due to the electron-phonon interaction otherwise the PIVC is achieved by the interaction of electrons with the photoinduced vibrations. It is worth to highlight that the latter case is the most probable.

## III. PHOTOINDUCED VIBRONIC STATES

Since the PIVC can be established by two ways depending on the energy separation between the DSs, we consider these two possibilities. Note that our consideration is close to the $E \otimes e$ problem [11] which is related to the RE ion doped crystals [19]. In contrast to the considered case, the Stark splitting of the electron levels of an RE ion is due to the action of the crystal field of the matrix rather than an electromagnetic field.

### A. Vibronic states due to the interaction of electrons with optical phonons

We use Eqs. (5) and (9) to determine the optical phonon contribution to the non-diagonal part of the electron-vibrational interaction in the DS basis. Finally, the interaction Hamiltonian takes the form

$$H' = \sum_{\mathbf{k},s,n} \hbar \left( G_0^{\mathbf{k},s} e^{-i2\Omega t} + G_0^{\mathbf{k},s*} e^{i2\Omega t} \right)$$
$$\times \left( \sqrt{n_{\mathbf{k},s}} e^{-i\omega_{\mathbf{k},s}t} + \sqrt{n_{\mathbf{k},s}+1} e^{i\omega_{\mathbf{k},s}t} \right),$$
$$G_0^{\mathbf{k},s} = U_m^+ U_m^- C_m^2 \left[ U_m^+ U_m^- \left( g_{11}^{\mathbf{k},s} - g_{22}^{\mathbf{k},s} \right) \right.$$
$$\left. + (U_m^+)^2 S_m g_{12}^{\mathbf{k},s} - (U_m^-)^2 S_m^* g_{21}^{\mathbf{k},s} \right]. \quad (13)$$

Then, we assume that there is a resonance between the transition frequency $2\Omega$ of the DSs and the optical phonon mode of frequency $\omega_{op}$. For each selected mode, the Hamiltonian of Eq. (13) can be treated as the Jaynes-Cummings interaction Hamiltonian. Transforming from matrix elements to the operators and introducing the parity operator, we get the interaction Hamiltonian in the following form [20]

$$\tilde{V}'_{op} = \exp\left\{ it\omega_{op}B^+B - it\Omega(-1)^{B^+B}\Pi \right\} \hbar G_0 \left( B + B^+ \right)$$
$$\times \exp\left\{ -it\omega_{op}B^+B + it\Omega(-1)^{B^+B}\Pi \right\}, \Pi = -\sigma_z(-1)^{b^+b},$$
$$B = \sigma_x b, \ B^+ = \sigma_x b^+. \quad (14)$$

Here, $\Pi$ is the parity operator. Since the Jaynes-Cummings Hamiltonian commute with the parity operator, the PIVSs in the DS basis can be written as

$$|v_{op}(t),m,n\rangle = D_n^+(t)|v_+,m,n\rangle + D_n^-(t)|v_-,m,n\rangle,$$
$$\Pi|v_+,m,n\rangle = |v_+,m,n\rangle, \ \Pi|v_-,m,n\rangle = -|v_-,m,n\rangle. \quad (15)$$

Here, signs + and – denote the even and odd parity, respectively. With use of the Schrödinger equation and the relations given in Appendix B, we get the following differential systems

$$\frac{\partial}{\partial t} D_n^+ = -iG_0 D_{n+1}^+ \sqrt{n+1} e^{-i\omega_{op}t} \exp\left\{ it2\Omega(-1)^{n+1} \right\}$$
$$-iG_0 D_{n-1}^+ \sqrt{n} e^{i\omega_{op}t} \exp\left\{ it2\Omega(-1)^{n-1} \right\},$$
$$\frac{\partial}{\partial t} D_n^- = -iG_0 D_{n+1}^- \sqrt{n+1} e^{-i\omega_{op}t} \exp\left\{ it2\Omega(-1)^{n} \right\}$$
$$-iG_0 D_{n-1}^- \sqrt{n} e^{i\omega_{op}t} \exp\left\{ it2\Omega(-1)^{n} \right\}. \quad (16)$$

Then, we assume the weak electron-phonon interaction that means using the RWA. Taking into account the parity of the numbers of phonon states, in the near-resonance conditions, Eqs. (16) transform into the following systems

$$\frac{\partial}{\partial t} D_{2n}^+(t) = -iG_0 \sqrt{2n} D_{2n-1}^+(t) e^{-i\delta_0 t},$$
$$\frac{\partial}{\partial t} D_{2n-1}^+(t) = -iG_0 \sqrt{2n} D_{2n}^+(t) e^{i\delta_0 t}, \quad (17)$$
$$\frac{\partial}{\partial t} D_{2n}^-(t) = -iG_0 \sqrt{2n+1} D_{2n+1}^-(t) e^{i\delta_0 t},$$
$$\frac{\partial}{\partial t} D_{2n+1}^-(t) = -iG_0 \sqrt{2n+1} D_{2n}^-(t) e^{-i\delta_0 t}, \quad (18)$$
$$\delta_0 = 2\Omega - \omega_{op}.$$

The solutions of Eqs. (17) and (18) for different initial conditions are well-known (for instance see Ref. [21]) but in the considered case these solutions are divided into two groups according to the parity of the phonon number at the DSs. Properly speaking, we should take the parity considerations into account in Sec. II where the DSs have been presented but we omit this fact for simplicity. Thus, the odd and even wave functions in the DS basis take the form

$$|v_-(t),m,n\rangle = D^-_{2n}(t)|\xi_1,m,2n\rangle + D^-_{2n+1}(t)|\xi_2,m,2n+1\rangle,$$
$$|v_+(t),m,n\rangle = D^+_{2n-1}(t)|\xi_1,m,2n-1\rangle + D^+_{2n}(t)|\xi_2,m,2n\rangle. \quad (19)$$

Moreover, these solutions can be written in three different bases (i) in the BS basis, (ii) in the DS basis, and (iii) in the basis of the doubly dressed states (DDSs). The DDSs are the states which are dressed first time by the electron-photon interaction and second time by the electron-phonon interaction. In the DDS basis, the parity wave functions read as

$$|v_\pm(t),m,n\rangle = Z_1^\pm|\zeta_1^\pm(t),m,n\rangle + Z_2^\pm|\zeta_2^\pm(t),m,n\rangle,$$
$$|\zeta_1^+(t),m,n\rangle = -s_n U^-_{2n}|\xi_1,m,2n-1\rangle e^{-iE_{11}^+ t}$$
$$+ U^+_{2n}|\xi_2,m,2n\rangle e^{-itE_{12}^+},$$
$$|\zeta_2^+(t),m,n\rangle = U^+_{2n}|\xi_1,m,2n-1\rangle e^{-iE_{21}^+ t}$$
$$+ s_n^* U^-_{2n}|\xi_2,m,2n\rangle e^{-iE_{22}^+ t},$$
$$|\zeta_1^-(t),m,n\rangle = U^+_{2n+1}|\xi_1,m,2n\rangle e^{-iE_{11}^- t}$$
$$- s_n^* U^-_{2n+1}|\xi_2,m,2n+1\rangle e^{-itE_{12}^-},$$
$$|\zeta_2^-(t),m,n\rangle = s_n U^+_{2n+1}|\xi_1,m,2n\rangle e^{-iE_{21}^- t}$$
$$+ U^-_{2n+1}|\xi_2,m,2n+1\rangle e^{-iE_{22}^- t},\ Z_1^+ = U^+_{2n}S_m\bar{U}_m^- - s_n^* U^-_{2n}\bar{U}_m^+,$$
$$Z_2^+ = U^+_{2n}\bar{U}_m^+ - s_n U^-_{2n}S_m\bar{U}_m^-,\ Z_1^- = U^+_{2n+1}\bar{U}_m^+ - s_n U^-_{2n+1}S_m\bar{U}_m^-,$$
$$Z_2^- = U^-_{2n+1}S_m\bar{U}_m^- + s_n^* U^+_{2n+1}\bar{U}_m^+,\ U_n^\pm = \sqrt{\frac{1}{2}\left(1\pm\frac{\delta_1}{2\Omega_n}\right)},$$
$$s_n = \frac{G_0}{|G_0|},\ 2\Omega_n = \sqrt{\delta_1^2 + 4n|G_0|^2},$$
$$E_{11}^\pm = -\Omega_{2n+0.5\mp 0.5} - 2\Omega + \left(2n\mp\frac{1}{2}\right)\omega_{op},$$
$$E_{12}^\pm = -\Omega_{2n+0.5\mp 0.5} + 2\Omega + \left(2n\mp\frac{1}{2}\right)\omega_{op},$$
$$E_{21}^\pm = \Omega_{2n+0.5\mp 0.5} - 2\Omega + \left(2n\mp\frac{1}{2}\right)\omega_{op},$$
$$E_{22}^\pm = \Omega_{2n+0.5\mp 0.5} + 2\Omega + \left(2n\mp\frac{1}{2}\right)\omega_{op}. \quad (20)$$

Eqs. (20) are written with the initial conditions that correspond with the electron population of the DSs in the electron-photon system [see Eq. (2)]. Additionally, we assume that the TLDS state is initially a product state which is composed by its subsystem states, namely $\bar{U}_m^+ = U_m^+ C_m D^\pm_{2n-0.5\mp 0.5}(0)$ and $\bar{U}_m^- = S_m U_m^- C_m D^\pm_{2n+0.5\mp 0.5}(0)$ where $C_m$ and $D^\pm_n(0)$ are the initial probability amplitudes for the photon and phonon states, respectively. Since all the DSs have the same energy of photon states, we omit the temporal dependencies of photon states to make Eqs. (20) less cumbersome.

### B. Vibronic states due to the interaction of electrons with forced vibrations

Using the interaction Hamiltonian of Eq. (12) and presenting the PIVSs as a linear combination of the DSs, we can define the coefficients of expansion from the following differential system

$$\frac{\partial}{\partial t}d_{1,n}(t) = -iF_0 e^{-i2\Omega t}d_{2,n}(t),$$
$$\frac{\partial}{\partial t}d_{2,n}(t) = -iF_0^* e^{i2\Omega t}d_{1,n}(t). \quad (21)$$

The solution of the system can be written in three different bases as it has been done in Sec. III A. In the considered case, the DDSs are formed with participation of the forced vibrations. Thus, the wave function takes the following form

$$|v_{fv}(t),m,n\rangle = \frac{1}{\sqrt{2}}|\zeta_1(t),m,n\rangle + \frac{1}{\sqrt{2}}|\zeta_2(t),m,n\rangle,$$
$$|\zeta_1(t),m,n\rangle = C_m U_m^+ W_1^+ e^{-i(2\Omega+\Omega_1)t}|\xi_1(t),m,n\rangle$$
$$+ C_m S_m U_m^- W_2^+ e^{i(2\Omega-\Omega_1)t}|\xi_2(t),m,n\rangle,$$
$$|\zeta_2(t),m,n\rangle = C_m U_m^+ W_1^- e^{-i(2\Omega-\Omega_1)t}|\xi_1(t),m,n\rangle$$
$$+ C_m S_m U_m^- W_2^- e^{i(2\Omega+\Omega_1)t}|\xi_2(t),m,n\rangle,$$
$$W_1^\pm = \frac{1}{\sqrt{2}} \pm \frac{U_m^+ F_0 - U_m^+\Omega}{\sqrt{2}U_m^+\Omega_1},\ W_2^\pm = \frac{1}{\sqrt{2}} \pm \frac{U_m^+ F_0^* + U_m^-\Omega}{\sqrt{2}U_m^-\Omega_1},$$
$$\Omega_1 = \sqrt{\Omega^2 + |F_0|^2}. \quad (22)$$

The initial conditions for the coefficients $d_{1,n}$ and $d_{2,n}$ are the same as in the case of the optical phonons (see Eqs. (20) and paragraph below). Eqs. (22) show that the electron population in the TLDS oscillates between the DSs with the frequency $\Omega_1$ and becomes equally divided between the DDSs. At the same time, the considered contribution of the electron-vibrational interaction does not describe the changing of the phonon states. In order to establish the coupling of the forced vibrations with the phonon modes, we must take into account the third term of the Taylor's series expansion in displacements for the interaction operator of Eq. (7). Using the wave functions of Eqs. (22), the non-diagonal part of the quadratic interaction can be written in the following form

$$H''_{nd} = \langle v_{fv}(t),m,n|\tilde{V}_{e-pn}|v_{fv}(t),m,n\rangle_{nd}$$
$$= \frac{C_m^2}{2}S_m U_m^- U_m^+ e^{i4\Omega t}\langle\xi_1(t),m,n|\tilde{V}_{e-pn}|\xi_2(t),m,n\rangle\left(W_1^{+*}W_2^+\right.$$
$$\left.+ W_1^{+*}W_2^- e^{i2\Omega_1 t} + W_1^{-*}W_2^+ e^{-i2\Omega_1 t} + W_1^{-*}W_2^-\right) + \text{H.c.} \quad (23)$$

Then, we transform the DS basis to the BS basis and get the expression containing the matrix elements of the electron-vibrational operator as it has been done before [see Eq. (5)]. The matrix elements of the electron-vibrational operator in the second order with respect to the displacements are defined in Appendix A. In the obtained expressions, we neglect the electron-phonon contribution and rewrite the EDEP contribution as

$$\Delta \tilde{V}_{22}'' = \sum_{\mathbf{k},s} \left(-\hbar^3\right) f_{\mathbf{k},s} h_{22}^{\mathbf{k},s} \left(\sqrt{n_{\mathbf{k},s}} e^{-i\omega_{\mathbf{k},s}t} + \sqrt{n_{\mathbf{k},s}+1} e^{i\omega_{\mathbf{k},s}t}\right),$$

$$\Delta \tilde{V}_{12}'' = \sum_{\mathbf{k},s} \hbar^3 f_{\mathbf{k},s} h_{12}^{\mathbf{k},s} e^{-i\delta t} \left[\left(n_{\mathbf{k},s}-\frac{1}{2}\right)\sqrt{n_{\mathbf{k},s}} e^{-i\omega_{\mathbf{k},s}t}\right.$$
$$\left. -\left(n_{\mathbf{k},s}+\frac{3}{2}\right)\sqrt{n_{\mathbf{k},s}+1} e^{i\omega_{\mathbf{k},s}t}\right],$$

$$\Delta \tilde{V}_{21}'' = \sum_{\mathbf{k},s} \left(-\hbar^3\right) f_{\mathbf{k},s} h_{21}^{\mathbf{k},s} e^{i\delta t} \left(n_{\mathbf{k},s}+\frac{1}{2}\right)\left[\sqrt{n_{\mathbf{k},s}} e^{-i\omega_{\mathbf{k},s}t}\right.$$
$$\left. -\sqrt{n_{\mathbf{k},s}+1} e^{i\omega_{\mathbf{k},s}t}\right]. \quad (24)$$

In this approximation, the non-diagonal part of the quadratic interaction can be readily found

$$H_{ED}'' = \frac{\hbar C_m^2}{2}\left[\left(W_1^{+*}W_2^+ + W_1^{-*}W_2^-\right)e^{i2\Omega t} + W_1^{+*}W_2^- e^{i2(\Omega+\Omega_1)t}\right.$$
$$\left. + W_1^{-*}W_2^+ e^{i2(\Omega-\Omega_1)t}\right]$$
$$\times \sum_{\mathbf{k},s,n} \left\{\left(F_{11}^{\mathbf{k},s}+F_2^{\mathbf{k},s}\right)\sqrt{n_{\mathbf{k},s}} e^{-i\omega_{\mathbf{k},s}t} + \left(F_{11}^{\mathbf{k},s}-F_2^{\mathbf{k},s}\right)\sqrt{n_{\mathbf{k},s}+1} e^{i\omega_{\mathbf{k},s}t}\right\}$$
$$+\frac{\hbar C_m^2}{2}\left[\left(W_2^{+*}W_1^+ + W_2^{-*}W_1^-\right)e^{-i2\Omega t} + W_2^{+*}W_1^- e^{-i2(\Omega+\Omega_1)t}\right.$$
$$\left. + W_2^{-*}W_1^+ e^{-i2(\Omega-\Omega_1)t}\right]$$
$$\times \sum_{\mathbf{k},s,n} \left\{\left(F_{12}^{\mathbf{k},s}-F_2^{\mathbf{k},s*}\right)\sqrt{n_{\mathbf{k},s}} e^{-i\omega_{\mathbf{k},s}t} + \left(F_{11}^{\mathbf{k},s}+F_2^{\mathbf{k},s*}\right)\sqrt{n_{\mathbf{k},s}+1} e^{i\omega_{\mathbf{k},s}t}\right\},$$

$$F_2^{\mathbf{k},s} = \hbar^2 f_{\mathbf{k},s} U_m^+ U_m^-\left(n_{\mathbf{k},s}+\frac{1}{2}\right)\left[(U_m^+)^2 S_m h_{12}^{\mathbf{k},s} + (U_m^-)^2 S_m^* h_{21}^{\mathbf{k},s}\right],$$

$$F_{11}^{\mathbf{k},s} = \hbar^2 f_{\mathbf{k},s} U_m^-(U_m^+)^2 \left(U_m^- h_{22}^{\mathbf{k},s} - U_m^+ S_m h_{12}^{\mathbf{k},s}\right),$$

$$F_{12}^{\mathbf{k},s} = \hbar^2 f_{\mathbf{k},s} U_m^+(U_m^-)^2 \left(U_m^+ h_{22}^{\mathbf{k},s} + U_m^- S_m h_{12}^{\mathbf{k},s}\right). \quad (25)$$

Here, $h_{ij}^{\mathbf{k},s}$ is the quadratic coupling strength of the electron-phonon interaction. It can be seen from Eq. (25) that the PIVS spectrum has three frequencies, namely, $2\Omega$, $2(\Omega - \Omega_1)$, and $2(\Omega + \Omega_1)$. Additionally, the diagonal matrix elements in the basis of the DSs give fourth frequency of $2\Omega_1$ (see Fig. 1 in Ref. [22]). This quartet resembles the Mollow triplet [23] in the electron-photon system in the conditions of the ac Stark effect.

Let us consider the term which defines the electron transition of frequency $2\Omega$. Other terms of $H_{ED}''$ can be accounted in the same manner. We assume that there is a resonance between the transition frequency $2\Omega$ of the DSs and acoustic phonon mode $s$ at the certain point $\mathbf{k}_0$. We also assume the weak electron-phonon coupling. These assumptions make it possible to use the RWA for the phonon mode. Finally, for each selected mode, $(s, \mathbf{k}_0)$, we get the similar Jaynes-Cummings interaction Hamiltonian as in Eq. (13). The final PIVSs in the DS basis take the form of Eq. (15) with the parity functions of Eq. (19). In the considered case, the expansion coefficients can be found from Eqs. (17) and (18) with substitutions $\delta_0 \to \delta_1$ and $G_0 \to G_{\mathbf{k}_0}$ where

$$\delta_1 = 2\Omega - \omega_{\mathbf{k}_0},$$
$$G_{\mathbf{k}_0} = \frac{C_m^2}{2}\left(F_{11}^{\mathbf{k}_0} + F_2^{\mathbf{k}_0}\right)\left(W_1^{+*}W_2^+ + W_1^{-*}W_2^-\right). \quad (26)$$

For an exact resonance, the coefficient module squares can be written as

$$\left|D_{2n-0.5\mp 0.5}^{\pm}(t)\right|^2 = \frac{1}{2} + \frac{(\bar{U}_m^+)^2-(\bar{U}_m^-)^2}{2}\cos\left(2\Omega_{20}^{\pm}t\right)$$
$$-\frac{\bar{U}_m^+\bar{U}_m^-}{2}i\left(S_m^*s_n^* - S_m s_n\right)\sin\left(2\Omega_{20}^{\pm}t\right),$$

$$\left|D_{2n+0.5\mp 0.5}^{\pm}(t)\right|^2 = \frac{1}{2} - \frac{(\bar{U}_m^+)^2-(\bar{U}_m^-)^2}{2}\cos\left(2\Omega_{20}^{\pm}t\right)$$
$$+\frac{\bar{U}_m^+\bar{U}_m^-}{2}i\left(S_m^*s_n^* - S_m s_n\right)\sin\left(2\Omega_{20}^{\pm}t\right),$$

$$\Omega_2^{\pm} = \sqrt{\delta_1^2 + 4\left(2n_{\mathbf{k}_0} + \frac{1}{2}\mp\frac{1}{2}\right)\left|G_{\mathbf{k}_0}\right|^2}. \quad (27)$$

Here, $\Omega_{20}$ is the resonance frequency with $\delta_1 = 0$. The quantities $\bar{U}_m^{\pm}$ are defined after Eqs. (20). Therefore, the final PIVSs of the corresponding parity can be presented in three different bases in the same manner as the wave functions of Eqs. (20). In the case of the DDS basis, the energy eigenvalues take the form similar as in the DS case

$$E_1^{\pm} = \omega_0(m+1) + \omega_{\mathbf{k}_0}\left(2n_{\mathbf{k}_0}\mp\frac{1}{2}\right) - \Omega_2^{\pm},$$

$$E_2^{\pm} = \omega_0(m+1) + \omega_{\mathbf{k}_0}\left(2n_{\mathbf{k}_0}\mp\frac{1}{2}\right) + \Omega_2^{\pm}. \quad (28)$$

Thus, the DDSs are formed by two steps in the case of the forced vibrations. At the first step, the basis of levels is created due to the first order electron-vibrational interaction. At the second step, the second order interaction leads to the broadening of the levels. Finally, taking into account the dispersion of acoustic phonon states, we obtain the quasi-continuous electron spectrum in the vicinity of the DSs [22].

## IV. CONCLUSIONS

We describe the mechanism of formation of the VC under the action of a strong electromagnetic field in the ac

Stark effect conditions for the TLDS. To this end, we include the electron-photon interaction in the zeroth-approximation and consider the electron-vibrational interaction in the basis of the DSs. In DS basis, the energy separation of electron levels is comparable with the vibrational energy of reservoir oscillators. As a result, the nonadiabatic mixing of the DSs occurs and the PIVSs appear. We estimate the contributions of different vibrational modes of reservoir oscillators into establishment of the PIVC. The strongest VC is achieved by the interaction of electrons with optical phonons in the case of the exact resonance conditions. If the resonance conditions for optical phonons are not satisfied, the VC is reached by the forced vibrations of reservoir oscillators. In this case, the PIVSs are formed by both the rapid changing of the electron density with the Rabi frequency and the coupling of the forced vibrations to phonon modes of the reservoir. Thus, the case of the forced vibrations is the most general. It worth to mention that the coupling of the forced vibrations to phonon modes can be used to control the energy transfer between the electron subsystem and the reservoir [22]. Finally, the obtained PIVSs possess the parity feature according to the phonon number of the DSs. Since the PIVSs are formed by both the electron-photon and electron-vibrational interactions, these states are DDSs. We obtain the expressions for the PIVSs in the case of the RWA. The use of the RWA means the weak electron-vibrational coupling in the TLDS. However, there are no limitations for defining these states in the case of the strong coupling. For this purpose, the generalized RWA may be used [20, 24].

## ACKNOWLEDGMENTS

Financial support of the Ministry of Education and Science of the Russian Federation (grant 074-U01) and of the Russian Foundation for Basic Research (grant 17-02-00598) are gratefully acknowledged.

## APPENDIX A: MATRIX ELEMENTS OF ELECTRON-VIBRATIONAL INTERACTION OPERATOR

The matrix element of the electron-vibrational operator that has the first order with respect to the displacements [see Eq. (7)] for the excited electron state can be readily obtained as

$$\tilde{V}_{22}^1 = \sum_{n,n'} \langle n' | \text{T}\exp\{i\hbar^{-1}\int dt' \Delta H(t')\} \sum_{\mathbf{k},s} \hbar g_{22}^{\mathbf{k},s} (\tilde{b}_{\mathbf{k},s} + \tilde{b}_{-\mathbf{k},s}^+)$$
$$\times \text{T}\exp\{-i\hbar^{-1}\int dt' \Delta H(t')\} | n \rangle$$
$$= \sum_{n,n'} \prod_{\mathbf{k}',s'} \langle n' | \{1 + \hbar f_{\mathbf{k}',s'}(\tilde{b}_{\mathbf{k}',s'} - \tilde{b}_{-\mathbf{k}',s'}^+) + \ldots\} \sum_{\mathbf{k},s} \hbar g_{22}^{\mathbf{k},s} (\tilde{b}_{\mathbf{k},s} + \tilde{b}_{-\mathbf{k},s}^+)$$
$$\times \prod_{\mathbf{k}'',s''} \{1 - \hbar f_{\mathbf{k}'',s''}(\tilde{b}_{\mathbf{k}'',s''} - \tilde{b}_{-\mathbf{k}'',s''}^+) + \ldots\} | n \rangle$$
$$\approx \sum_{\mathbf{k},s,n,n'} \hbar g_{22}^{\mathbf{k},s} \langle n' | (\tilde{b}_{\mathbf{k},s} + \tilde{b}_{-\mathbf{k},s}^+) | n \rangle$$
$$+ \sum_{n,n'} \sum_{\mathbf{k},s,\mathbf{k}',s'} \hbar^2 g_{22}^{\mathbf{k},s} f_{\mathbf{k}',s'} \langle n' | (\tilde{b}_{\mathbf{k}',s'} - \tilde{b}_{-\mathbf{k}',s'}^+)(\tilde{b}_{\mathbf{k},s} + \tilde{b}_{-\mathbf{k},s}^+) | n \rangle$$
$$+ \sum_{n,n'} \sum_{\mathbf{k},s,\mathbf{k}'',s''} \hbar^2 g_{22}^{\mathbf{k},s} f_{\mathbf{k}'',s''} \langle n' | (\tilde{b}_{\mathbf{k},s} + \tilde{b}_{-\mathbf{k},s}^+)(\tilde{b}_{-\mathbf{k}'',s''}^+ - \tilde{b}_{\mathbf{k}'',s''}) | n \rangle$$
$$= \sum_{\mathbf{k},s,n} \hbar g_{22}^{\mathbf{k},s} \left(\sqrt{n_{\mathbf{k},s}} e^{-i\omega_{\mathbf{k},s}t} + \sqrt{n_{\mathbf{k},s}+1} e^{i\omega_{\mathbf{k},s}t}\right)$$
$$+ \sum_{\mathbf{k},s} 2\hbar^2 g_{22}^{\mathbf{k},s} f_{\mathbf{k},s}. \quad (A1)$$

In Eq. (A1), we use the series expansion of the Texp functions and leave the terms up to the first order with respect to the displacement operator, $\boldsymbol{\Lambda}$. The non-diagonal matrix elements are determined in the same manner as

$$\tilde{V}_{12}^1 = \sum_{\mathbf{k},s,n,n'} \hbar g_{12}^{\mathbf{k},s} e^{-i\delta t} \langle n' | (\tilde{b}_{\mathbf{k},s} + \tilde{b}_{-\mathbf{k},s}^+) \text{T}\exp\{-i\hbar^{-1}\int dt' \Delta H(t')\} | n \rangle$$
$$= \sum_{\mathbf{k},s,n,n'} \hbar g_{12}^{\mathbf{k},s} e^{-i\delta t} \langle n' | (\tilde{b}_{\mathbf{k},s} + \tilde{b}_{-\mathbf{k},s}^+)$$
$$\times \prod_{\mathbf{k}'',s''} \{1 - \hbar f_{\mathbf{k}'',s''}(\tilde{b}_{\mathbf{k}'',s''} - \tilde{b}_{-\mathbf{k}'',s''}^+) + \ldots\} | n \rangle$$
$$\approx \sum_{\mathbf{k},s,n,n'} \hbar g_{12}^{\mathbf{k},s} e^{-i\delta t} \langle n' | (\tilde{b}_{\mathbf{k},s} + \tilde{b}_{-\mathbf{k},s}^+) | n \rangle$$
$$+ \sum_{n,n'} \sum_{\mathbf{k},s,\mathbf{k}'',s''} \hbar^2 g_{12}^{\mathbf{k},s} f_{\mathbf{k}'',s''} e^{-i\delta t} \langle n' | (\tilde{b}_{\mathbf{k},s} + \tilde{b}_{-\mathbf{k},s}^+)(\tilde{b}_{-\mathbf{k}'',s''}^+ - \tilde{b}_{\mathbf{k}'',s''}) | n \rangle$$
$$= \sum_{\mathbf{k},s,n} \hbar g_{12}^{\mathbf{k},s} e^{-i\delta t} \left(\sqrt{n_{\mathbf{k},s}} e^{-i\omega_{\mathbf{k},s}t} + \sqrt{n_{\mathbf{k},s}+1} e^{i\omega_{\mathbf{k},s}t}\right)$$
$$+ \sum_{\mathbf{k},s} \hbar^2 g_{12}^{\mathbf{k},s} e^{-i\delta t} f_{\mathbf{k},s}, \quad (A2)$$

and analogously we find

$$\tilde{V}_{21}^1 = \sum_{\mathbf{k},s,n} \hbar g_{21}^{\mathbf{k},s} e^{i\delta t} \left(\sqrt{n_{\mathbf{k},s}} e^{-i\omega_{\mathbf{k},s}t} + \sqrt{n_{\mathbf{k},s}+1} e^{i\omega_{\mathbf{k},s}t}\right)$$
$$+ \sum_{\mathbf{k},s} \hbar^2 g_{21}^{\mathbf{k},s} e^{i\delta t} f_{\mathbf{k},s}. \quad (A3)$$

Dividing these matrix elements into two parts, we can rewrite Eqs. (A1)-(A3) as a sum of two contributions that describe the electron-phonon interaction and the interaction of electrons and the forced vibrations, respectively. Thus, we get Eqs. (11).

The matrix elements of the second order with respect to the displacements [third term of Eq. (7)] can be written as

$$\tilde{V}_{22}^2 = \sum_{n,n'} \langle n' | \text{T}\exp\{i\hbar^{-1}\int dt' \Delta H(t')\}$$
$$\times \sum_{\mathbf{k},s} (-\hbar^2) h_{22}^{\mathbf{k},s} \left(\tilde{b}_{\mathbf{k},s}^+ \tilde{b}_{\mathbf{k},s} + \frac{1}{2}\right) \text{T}\exp\{-i\hbar^{-1}\int dt' \Delta H(t')\} | n \rangle$$
$$\approx \sum_{\mathbf{k},s,n,n'} (-\hbar^2) h_{22}^{\mathbf{k},s} \langle n' | \left(\tilde{b}_{\mathbf{k},s}^+ \tilde{b}_{\mathbf{k},s} + \frac{1}{2}\right) | n \rangle$$
$$+ \sum_{n,n'} \sum_{\mathbf{k},s,\mathbf{k}',s'} \hbar^3 h_{22}^{\mathbf{k},s} f_{\mathbf{k}',s'} \langle n' | (\tilde{b}_{-\mathbf{k}',s'}^+ - \tilde{b}_{\mathbf{k}',s'})\left(\tilde{b}_{\mathbf{k},s}^+ \tilde{b}_{\mathbf{k},s} + \frac{1}{2}\right)$$

$$-\sum_{n,n'}\sum_{\mathbf{k},s,\mathbf{k}'',s''}\hbar^3 h_{22}^{\mathbf{k},s} f_{\mathbf{k}'',s''}\langle n'|\left(\tilde{b}_{\mathbf{k},s}^+\tilde{b}_{\mathbf{k},s}+\frac{1}{2}\right)\left(\tilde{b}_{-\mathbf{k}'',s''}^+-\tilde{b}_{\mathbf{k}'',s''}\right)|n\rangle$$

$$=\sum_{\mathbf{k},s,n}\left(-\hbar^2\right)h_{22}^{\mathbf{k},s}\left(n_{\mathbf{k},s}+\frac{1}{2}\right)$$

$$+\sum_{\mathbf{k},s,n}\left(-\hbar^3\right)h_{22}^{\mathbf{k},s}f_{\mathbf{k},s}\left(\sqrt{n_{\mathbf{k},s}}e^{-i\omega_{\mathbf{k},s}t}+\sqrt{n_{\mathbf{k},s}+1}e^{i\omega_{\mathbf{k},s}t}\right), \quad (A4)$$

$$\tilde{V}_{12}^2=\sum_{\mathbf{k},s,n,n'}\hbar^2 h_{12}^{\mathbf{k},s}e^{-i\delta t}\langle n'|\left(\tilde{b}_{\mathbf{k},s}^+\tilde{b}_{\mathbf{k},s}+\frac{1}{2}\right)$$
$$\times\mathrm{T}\exp\left\{-i\hbar^{-1}\int dt'\Delta H(t')\right\}|n\rangle$$

$$\approx\sum_{\mathbf{k},s,n,n'}\left(-\hbar^2\right)h_{22}^{\mathbf{k},s}e^{-i\delta t}\langle n'|\left(\tilde{b}_{\mathbf{k},s}^+\tilde{b}_{\mathbf{k},s}+\frac{1}{2}\right)|n\rangle$$

$$-\sum_{n,n'}\sum_{\mathbf{k},s,\mathbf{k}'',s''}\hbar^3 h_{12}^{\mathbf{k},s}e^{-i\delta t}f_{\mathbf{k}'',s''}\langle n'|\left(\tilde{b}_{\mathbf{k},s}^+\tilde{b}_{\mathbf{k},s}+\frac{1}{2}\right)\left(\tilde{b}_{-\mathbf{k}'',s''}^+-\tilde{b}_{\mathbf{k}'',s''}\right)|n\rangle$$

$$=\sum_{\mathbf{k},s,n}\left(-\hbar^2\right)h_{12}^{\mathbf{k},s}e^{-i\delta t}\left(n_{\mathbf{k},s}+\frac{1}{2}\right)$$

$$+\sum_{\mathbf{k},s,n}\hbar^3 h_{12}^{\mathbf{k},s}e^{-i\delta t}f_{\mathbf{k},s}\left[\left(n_{\mathbf{k},s}-\frac{1}{2}\right)\sqrt{n_{\mathbf{k},s}}e^{-i\omega_{\mathbf{k},s}t}\right.$$
$$\left.-\left(n_{\mathbf{k},s}+\frac{3}{2}\right)\sqrt{n_{\mathbf{k},s}+1}e^{i\omega_{\mathbf{k},s}t}\right], \quad (A5)$$

$$\tilde{V}_{21}^2=\sum_{\mathbf{k},s,n}\left(-\hbar^2\right)h_{21}^{\mathbf{k},s}e^{i\delta t}\left(n_{\mathbf{k},s}+\frac{1}{2}\right)$$

$$-\sum_{\mathbf{k},s,n}\hbar^3 h_{12}^{\mathbf{k},s}e^{i\delta t}f_{\mathbf{k},s}\left[\left(n_{\mathbf{k},s}+\frac{1}{2}\right)\sqrt{n_{\mathbf{k},s}}e^{-i\omega_{\mathbf{k},s}t}\right.$$
$$\left.-\left(n_{\mathbf{k},s}+\frac{3}{2}\right)\sqrt{n_{\mathbf{k},s}+1}e^{i\omega_{\mathbf{k},s}t}\right], \quad (A6)$$

$$h_{ij}^{\mathbf{k},s}=\sum_{\alpha}\frac{2k^2 v(2\mathbf{k})}{Nm_\alpha\hbar\omega_{\mathbf{k},s}}\langle\varphi_i|e^{i2\mathbf{k}\cdot\mathbf{r}}|\varphi_j\rangle. \quad (A7)$$

In Eqs. (A4)-(A7), the contributions due to the interaction of electrons and the forced vibrations contain the quantity $f_{\mathbf{k},s}$. These contributions are written in Eqs. (24).

## APPENDIX B: ELECTRON-PHONON INTERACTION WITH PARITY OPERATOR

The electron-phonon interaction operator in the interaction picture, containing the parity operator [see Eq. (14)], can be represented utilizing the following relations for the phonon and electron operators

$$\exp\left\{-it\Omega(-1)^{b^+ b}\Pi\right\}b\exp\left\{it\Omega(-1)^{b^+ b}\Pi\right\}$$
$$=b\exp\left\{it\Omega\left[(-1)^{b^+ b}-(-1)^{b^+ b-1}\right]\Pi\right\},$$

$$\exp\left\{-it\Omega(-1)^{b^+ b}\Pi\right\}b^+\exp\left\{it\Omega(-1)^{b^+ b}\Pi\right\}$$
$$=b^+\exp\left\{it\Omega\left[(-1)^{b^+ b}-(-1)^{b^+ b+1}\right]\Pi\right\},$$

$$\exp\left\{-it\Omega(-1)^{b^+ b}\Pi\right\}\sigma_x\exp\left\{it\Omega(-1)^{b^+ b}\Pi\right\}$$
$$=\sigma_x\exp\left\{it2\Omega(-1)^{b^+ b}\Pi\right\}. \quad (B1)$$

In Eq. (B1), we write only relations which contain the parity operator. Other relations can be written in the usual forms (see Ref. [21]). Therefore, the electron-phonon interaction operator reads as

$$\tilde{V}'_{\mathrm{op}}=\hbar G_0 e^{-i\omega_{\mathrm{op}}t}B\exp\left\{it\Omega\left[(-1)^{B^+ B}-(-1)^{B^+ B-1}\right]\Pi\right\}$$
$$+\hbar G_0 e^{i\omega_{\mathrm{op}}t}B^+\exp\left\{it\Omega\left[(-1)^{B^+ B}-(-1)^{B^+ B+1}\right]\Pi\right\}. \quad (B2)$$

To determine the action of the exponential operators in Eq. (B2), we expand the exponential function in series and take into account the parity of the states. Thus, we get the following relations

$$e^{-i\omega_{\mathrm{op}}t}B\exp\left\{it2\Omega(-1)^{B^+ B}\Pi\right\}|v_+(t),m,n\rangle$$
$$=e^{-i\omega_{\mathrm{op}}t}\sqrt{n}\exp\left\{it2\Omega(-1)^n\right\}|v_+(t),m,n-1\rangle,$$

$$e^{-i\omega_{\mathrm{op}}t}B\exp\left\{it2\Omega(-1)^{B^+ B}\Pi\right\}|v_-(t),m,n\rangle$$
$$=e^{-i\omega_{\mathrm{op}}t}\sqrt{n}\exp\left\{it2\Omega(-1)^{n-1}\right\}|v_-(t),m,n-1\rangle,$$

$$e^{i\omega_{\mathrm{op}}t}B^+\exp\left\{it2\Omega(-1)^{B^+ B}\Pi\right\}|v_+(t),m,n\rangle$$
$$=e^{i\omega_{\mathrm{op}}t}\sqrt{n+1}\exp\left\{it2\Omega(-1)^n\right\}|v_+(t),m,n+1\rangle,$$

$$e^{i\omega_{\mathrm{op}}t}B^+\exp\left\{it2\Omega(-1)^{B^+ B}\Pi\right\}|v_-(t),m,n\rangle$$
$$=e^{i\omega_{\mathrm{op}}t}\sqrt{n+1}\exp\left\{it2\Omega(-1)^{n+1}\right\}|v_-(t),m,n+1\rangle. \quad (B3)$$

The obtained relations are used in the Schrödinger equation to derive Eqs. (16).

[1] M. Atanasov, C. Daul, P. L. W. Tregenna-Piggott (eds), *Vibronic Interactions and the Jahn-Teller Effect: Theory and Applications* (Springer, Berlin, 2012).
[2] S. Weiler, A. Ulhaq, S. M. Ulrich, D. Richter, M. Jetter, P. Michler, C. Roy, and S. Hughes, Phonon-assisted incoherent excitation of a quantum dot and its emission properties, Phys. Rev. B **86**, 241304(R) (2012).
[3] A. S. Davydov, *Theory of Molecular Excitons* (Plenum, New-York, 1971).
[4] *The Dynamical Jahn–Teller Effect in Localized Systems*, edited by Yu.E. Perlin and M. Wagner (Elsevier, Amsterdam, 1984).

[5] V. M. Huxter, T. A. A. Oliver, D. Budker, and G. R. Fleming, Vibrational and electronic dynamics of nitrogen–vacancy centres in diamond revealed by two-dimensional ultrafast spectroscopy, Nature Phys. **9**, 744 (2013).

[6] A Lupei, V Lupei, C Presura, V N Enaki and A Petraru, Electron-phonon coupling effects on $Yb^{3+}$ spectra in several laser crystals, J. Phys.: Condens. Matter **11**, 3769 (1999).

[7] R. Schirhagl, K. Chang, M. Loretz, and C. L. Degen, "Nitrogen vacancy centers in diamond: nanoscale sensors for physics and biology," Ann. Rev. Phys. Chem. **65**, 83 (2014).

[8] A. Mysyrowicz, D. Hulin, A. Antonetti, A. Migus, W. T. Masselink, and H. Morkoç, "Dressed Excitons" in a Multiple-Quantum-Well Structure: Evidence for an Optical Stark Effect with Femtosecond Response Time, Phys. Rev. Lett. **56**, 2748 (1986).

[9] M. A. Bondarev, E. Yu. Perlin, and A. V. Ivanov, Multiphoton Absorption Controlled by the Resonance Optical Stark Effect in Crystals, Opt. and Spectr. **115**, 827 (2013).

[10] Y. Kayanuma and K. G. Nakamura, Dynamic Jahn-Teller viewpoint for generation mechanism of asymmetric modes of coherent phonons, Phys. Rev. B **95**, 104302 (2017).

[11] I. B. Bersuker, V. Z. Polinger, *Vibronic Interactions in Molecules and Crystals* (Springer, Berlin, 1989).

[12] H. J. Zeiger, J. Vidal, T. K. Cheng, E. P. Ippen, G. Dresselhaus, and M. S. Dresselhaus, Phys. Rev. B **45**, 768 (1992).

[13] A. V. Kuznetsov and C. J. Stanton, Theory of Coherent Phonon Oscillations in Semiconductors, Phys. Rev. Lett. **73**, 3243 (1994).

[14] U. Öpik, and M. H. L. Pryce, Studies of the Jahn Teller Effect. 1. A Survey of the Static Problem, Proc. R. Soc. A. **238**, 425 (1957).

[15] A. J. Brash, L. M. P. P. Martins, A. M. Barth, F. Liu, J. H. Quilter, M. Glässl, V. M. Axt, A. J. Ramsay, M. S. Skolnick, and A. M. Fox, Dynamic vibronic coupling in InGaAs quantum dots, JOSA B **33**, C115 (2016).

[16] N. B. Delone, V.P. Krainov, *Atoms in Strong Light Fields* (Springer, Berlin, 1985).

[17] Yu. E. Perlin, Modern methods in the theory of many-phonon processes, Sov. Phys. Usp. **6**, 542 (1964).

[18] R. Feynman, An Operator Calculus Having Applications in Quantum Electrodynamics, Phys. Rev. **84**, 108 (1951).

[19] T. Vaikjärv, V. Hizhnyakov, Time-dependent pseudo Jahn-Teller effect: Phonon-mediated long-time nonadiabatic relaxation, J. Chem. Phys. **140**, 064105 (2014).

[20] J. Casanova, G. Romero, I. Lizuain, J. J. Garcia-Ripoll, and E. Solano, Deep Strong Coupling Regime of the Jaynes-Cummings Model, Phys. Rev. Lett. **105**, 263603 (2010).

[21] M. Scully and M. Zubairy, Quantum optics (New York: Cambridge University Press, 1997).

[22] A. V. Ivanov, Energy Transfer Controlled by Dynamical Stark Shift in Two-level Dissipative Systems, Phys. Rev. Lett.

[23] B. R. Mollow, Power Spectrum of Light Scattered by Two-Level Systems, Phys. Rev. **188**, 1969 (1969).

[24] E. K. Irish, Generalized Rotating-Wave Approximation for Arbitrarily Large Coupling, Phys. Rev. Lett. **99**, 173601 (2007).